\documentclass[aps,prd,superscriptaddress,onecolumn,showpacs,nofootinbib,preprintnumbers]{revtex4}

\usepackage{amsmath,amssymb,hyperref,subfigure,xspace}
\usepackage{mciteplus,color,mathtools,graphicx}
\usepackage[english]{babel}

\newcommand{\ls}{LS\xspace}
\newcommand{\cf}{{\it cf.}\xspace}
\newcommand{\cpt}{CPM\xspace}
\newcommand{\diff}{\textrm{d}}
\newcommand{\ket}[1]{\left|#1\right\rangle}
\newcommand{\braket}[2]{\left\langle#1\right.\left|#2\right\rangle}
\newcommand{\sw}{$S$-}
\newcommand{\pw}{$P$-}
\newcommand{\dw}{$D$-}
\newcommand{\babar}{BaBar}
\newcommand{\belle}{Belle\xspace}
\newcommand{\lhcb}{LHCb\xspace}
\newcommand{\aka}{{\it aka}\xspace}
\newcommand{\eg}{{\it e.g.}\xspace}
\newcommand{\ie}{{\it i.e.}\xspace}
\newcommand{\coupl}[3]{%
{\ensuremath{F^{(#2),#3}_{#1}}}}

\begin{document}

\title{What is the right formalism to search for resonances?}

\author{M.~Mikhasenko}
\email{mikhail.mikhasenko@hiskp.uni-bonn.de}
\affiliation{{Universit\"at Bonn,
Helmholtz-Institut f\"ur Strahlen- und Kernphysik, 53115 Bonn, Germany}}

\author{A.~Pilloni}
\email{pillaus@jlab.org}
\affiliation{Theory Center, Thomas Jefferson National Accelerator Facility,
Newport News, VA 23606, USA}

\author{J.~Nys}
\affiliation{Theory Center, Thomas Jefferson National Accelerator Facility,
Newport News, VA 23606, USA}
\affiliation{Department of Physics and Astronomy, Ghent University, Belgium}
\affiliation{Center for Exploration of Energy and Matter, Indiana University, Bloomington, IN 47403, USA}
\affiliation{Physics Department, Indiana University, Bloomington, IN 47405, USA}

\author{M.~Albaladejo}
\affiliation{Departamento de F\'isica, Universidad de Murcia, E-30071 Murcia, Spain}

\author{C.~Fern\'andez-Ram\'irez}
\affiliation{Instituto de Ciencias Nucleares,
Universidad Nacional Aut\'onoma de M\'exico, Ciudad de M\'exico 04510, Mexico}

\author{A.~Jackura}
\affiliation{Center for Exploration of Energy and Matter,
Indiana University, Bloomington, IN 47403, USA}
\affiliation{Physics Department, Indiana University, Bloomington, IN 47405, USA}

\author{V.~Mathieu}
\affiliation{Theory Center, Thomas Jefferson National Accelerator Facility,
Newport News, VA 23606, USA}

\author{N.~Sherrill}
\affiliation{Center for Exploration of Energy and Matter,
Indiana University, Bloomington, IN 47403, USA}
\affiliation{Physics Department, Indiana University, Bloomington, IN 47405, USA}

\author{T.~Skwarnicki}
\affiliation{Syracuse University, Syracuse, NY 13244, USA}

\author{A.~P.~Szczepaniak}
\affiliation{Theory Center, Thomas Jefferson National Accelerator Facility,
Newport News, VA 23606, USA}
\affiliation{Center for Exploration of Energy and Matter,
Indiana University, Bloomington, IN 47403, USA}
\affiliation{Physics Department, Indiana University, Bloomington, IN 47405, USA}

\collaboration{Joint Physics Analysis Center}

\preprint{JLAB-THY-17-2606}


\begin{abstract}
Hadron decay chains constitute
one of the main sources of information on the QCD spectrum.
We discuss the differences between several partial wave analysis formalisms used in the literature to build the amplitudes.
We match the helicity amplitudes to the covariant tensor basis. Hereby, we pay attention to the analytical properties of the amplitudes and separate singularities of kinematical and dynamical nature.
We study the analytical properties of the spin-orbit (\ls) formalism, and some of the covariant tensor approaches.
In particular, we explicitly build the amplitudes for the
$B\to\psi \pi K$ and
$B\to\bar{D}\pi\pi$
decays,
and show that the energy dependence of the covariant approach is model dependent. We also show that the usual recursive construction of covariant tensors
explicitly violates crossing symmetry, which would lead to different resonance parameters extracted from scattering and decay processes.
\end{abstract}

\maketitle

\section{Introduction}
\label{sec:intro}
The high quality data on hadron production and decays that are or will be
available from \babar, BelleII, BESIII, CMS, CLAS12, COMPASS, GlueX, LHCb, and other experiments, necessitate rigorous amplitude analysis.
This is particularly true for the extraction of resonance parameters that are based on analytical partial waves.
Moreover, analytical reaction amplitudes are needed in conjunction with lattice data to study the hadron spectrum from first-principles lattice QCD calculations \cite{Wilson:2015dqa,Briceno:2016mjc,Briceno:2017qmb,Briceno:2017max,Hu:2016shf}.

In this paper, we focus on three-body decays, \aka 1-to-3 processes. In recent years such reactions have led to ample data that
resulted in the observation of new exotic phenomena, \eg the so-called XYZ states in heavy meson decays~\cite{Lebed:2016hpi,Esposito:2016noz,Olsen:2017bmm}, and that are also used in studies of excited mesons and baryons.
The issues we address and the methodology we present are, however, of relevance to other analyses as well, for example to baryon resonance studies in photoproduction \cite{Anisovich:2011fc,Workman:2012jf},
or meson spectroscopy from pion or photon beam fragmentation \cite{Abbon:2014aex,Shepherd:2009zz,Glazier:2015cpa}.

In the modern literature, there seems to be a lot of confusion regarding properties of the reaction amplitudes employed in analyses of such processes.
This is often stated in the context of a potentially nonrelativistic character
of certain approaches~\cite{Chung:1993da,Anisovich:2011fc,Adolph:2015tqa}.
As we explain below, however, rather than arising from relativistic kinematics, the differences between the various formalisms have a dynamical origin.
Reaction amplitudes are given by the scattering matrix elements between initial and final states that represent asymptotically free particles. Such states belong to a unitary, noncovariant representation of the Lorentz group. Since the scattering operator is a Lorentz scalar, reaction amplitudes share the transformation properties of the free particle states. A typical three particle decay process is dominated by \mbox{two-body} resonances, and can be well approximated by a finite number of partial waves. The latter can be given by the helicity partial waves or the Russell-Saunders, \aka \ls amplitudes~\cite{Collins:1977jy}.
For the \ls amplitudes, one couples particle states in the canonical representation. The relation between the helicity and \ls basis is a straightforward orthogonal transformation.
Because of the noncovariant transformation properties of the reaction amplitude, partial waves transform in a nontrivial way as well, \eg helicity amplitudes mix under Lorentz boosts through Wigner rotations. Nevertheless, all of the amplitudes referred to above (the helicity amplitudes, the helicity partial waves, the \ls partial wave amplitudes) are {\it relativistic}, \ie have well defined behavior under Lorentz transformations.

Since the helicity amplitudes involve asymptotically free particle states, they must be proportional to free particle wave functions, \eg Dirac spinors or polarization tensors.
These wave functions have mixed transformation properties, \ie have both covariant (Lorentz or Dirac), and noncovariant (helicity) indices.
The Lorentz and Dirac indices  need to be contracted with covariant tensors built from particle four-vectors and Dirac gamma matrices to yield the noncovariant helicity amplitudes. Helicity amplitudes can therefore be expressed as linear combinations of products of covariant tensors and wave functions with coefficients that are scalar functions of the Mandelstam invariants.
It can be shown that these scalar functions have only dynamical singularities as demanded by unitarity~\cite{Cohen-Tannoudji:1968lnm}, and for this reason are useful when analyzing singularities of the partial waves. Furthermore, these scalar functions are invariant under crossing which makes them convenient to relate amplitudes in the decay and scattering kinematics.

There exist an approach for constructing the scalar functions from an assumed model for the partial waves, hereafter referred to as the covariant projection method (\cpt)~\cite{Chung:1993da,Chung:2007nn,Filippini:1995yc,Anisovich:2006bc},
that starts from a \ls partial wave model (or equivalently the  Cartesian, \aka Zemach amplitudes~\cite{Zemach:1968zz}) but writes them
 in a covariant fashion.
The method has a drawback, which is related to the  behavior under crossing (see Section~\ref{sec:BDpipi}). The alternative, which we refer to as
the canonical approach~\cite{Jacob:1959at,Chung:1971ri,Collins:1977jy,cookbook}, is to use the well known relation between the helicity amplitudes and the helicity partial waves~\cite{Collins:1977jy} to determine the scalar functions in terms of the partial wave models.
The differences between these two approaches to relate partial waves and scalar functions result in factors which are confusingly referred in the literature as ``relativistic corrections''. These are actually Lorentz invariant functions and therefore can be absorbed into the scalar functions. In both the \cpt and canonical approaches, the relativistic kinematics is properly taken into account.
Thus, the differences in these approaches are dynamical in nature.

In what follows, we present a detailed comparison of these two approaches, paying specific attention to the analytical properties, which
are among the few constraints that
can be imposed in a model independent way. Instead of presenting results for a general case, we find it more pedagogical to compare these constructions in a few concrete examples.
The examples we discuss are of special interest to various ongoing analyses, and are complex enough to illustrate the general principles.
The first example is the parity violating (PV) three-body decay $B^0 \to \psi \pi^- K^+$, with $\psi = J/\psi, \psi(2S)$.
The analyses by Belle and LHCb show nontrivial structures appearing in the $\psi(2S) \,\pi$~\cite{Mizuk:2009da,Chilikin:2013tch,Aaij:2015wza,Aaij:2015zxa}, and in the $J/\psi \,\pi$ channel~\cite{Chilikin:2014bkk}.
These are of particular interest, because a resonance in these channels would require an exotic interpretation~\cite{Lebed:2016hpi,Esposito:2016noz,Olsen:2017bmm}.
The rest of the paper is organized as follows. In Sec.~\ref{sec:s.channel} we discuss the canonical approach on the example of the  $B\to \psi\pi K$ decay.  By relating the helicity partial waves to the scalar amplitudes via the partial wave expansion, we derive constraints and isolate the kinematical singularities. We also discuss implication of these constraints for the \ls partial wave amplitudes. The details of the amplitude parameterizations are given in the Appendices and are presented in a way that can be implemented in the standard data analysis tools~\cite{amptools,rootpwa}.
In Sec.~\ref{sec:comparison} we examine the \cpt approach and compare this model with the findings from Sec.~\ref{sec:s.channel}. We mention the crossing symmetry properties of \cpt using, as an example, $B^0 \to \bar D^0 \pi^+ \pi^-$, which was recently analyzed by LHCb within this formalism~\cite{Aaij:2015sqa}.
Summary and conclusions are given in Section~\ref{sec:sc}.

\section{Analyticity constraints for \texorpdfstring{$B\to \psi \pi K$}{B -> psi pi K}}
\label{sec:s.channel}

\begin{figure}[t]
\centering
\subfigure[\ Decay]{
\includegraphics[]{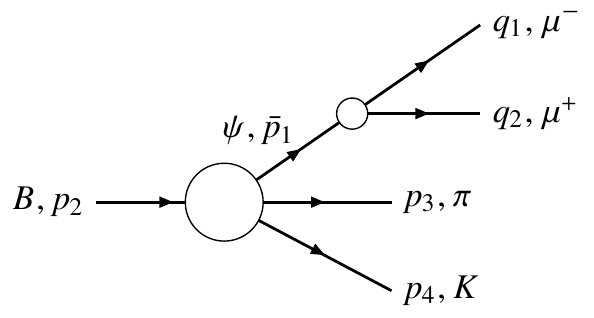} \label{fig:decay.diag1}}
\hspace{2cm}
\subfigure[\ $s$-channel scattering]{
\includegraphics[]{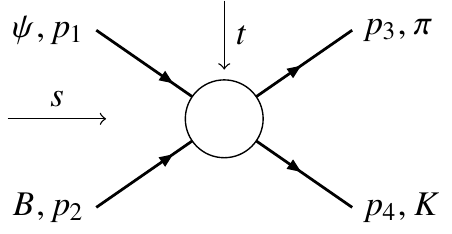}\label{fig:decay.diag2}}
\caption{
Reaction diagrams for (a) the $B\to \psi(\to\mu^-\mu^+) \pi K$ decay process, and for (b)
 the $\psi B\to \pi K$ $s$-channel scattering process. The $t$-channel process $\psi \pi\to \bar B K$  is indicated by the vertical line.} \label{fig:decay.diag}
\end{figure}

In Fig.~\ref{fig:decay.diag} we show
a diagram representing the kinematics of the decay $B \to \psi(\to \mu^+\mu^-) \pi K$. The spinless particles $B$, $\pi$, $K$ are stable against the strong interaction. The $\psi$ is narrow enough
to completely factorize its decay dynamics. Thus, we construct the amplitude considering $\psi$ to be stable. More details, including the dilepton decay of the $\psi$, are given in
Appendices~\ref{sec:parity.conserving} and~\ref{sec:our.model}.
We use $p_i$, $i=2\dots 4$ to label the momenta of $B$, $\pi$, and $K$ respectively. The momentum of the $\psi$ will be denoted by $\bar p_1$, for a reason which we will explain below.
The helicity amplitude for the decay process $B\to \psi \pi K$ is denoted by $\mathcal{A}_\lambda(s,t)$,
$\lambda$ being the helicity of $\psi$,
\ie \mbox{$\langle \psi\pi K, \text{out}| B,  \text{in}\rangle = (2\pi)^4\delta^4(p_2-\bar p_1-p_3-p_4) \mathcal{A}_\lambda$.}
The amplitude depends on the standard Mandelstam variables $s = (p_3+p_4)^2$, $t = (\bar p_1+p_3)^2$, and $u = (\bar p_1 + p_4)^{2}$ with $s + t + u = \sum_i m_i^2$.

The $B$ meson decays weakly, so $\mathcal{A}_\lambda$
is given by the sum of a PV and parity conserving (PC) amplitude.
The difficulty treating the decay channel directly is that
the mass of the decaying particle 
should be considered on the same footing 
as the other dynamical variables ($s$, $t$, $u$). This is demanded by unitarity, which implies that above a threshold,
the amplitude is a singular function of the corresponding dynamical variable. It is therefore simpler to study singularities
in a scattering channel and cross to the other channels by analytical continuation
in the momentum of the $\psi$, \ie by setting $\bar{p}_1 = -p_1$~\cite{Trueman:1964zzb}.
In general, under crossing, helicity amplitudes are mixed by Wigner rotations.
In our case, however, since crossing can be realized through a (unphysical)
boost in the direction of motion of the $\psi$,
there is no change in helicity.

We begin with the discussion of the PV amplitudes in the $s$-channel. The $s$-channel resonances correspond to the $K^*$'s and dominate the reaction. As discussed in the previous section, the analysis of the experimental data indicates a possible signal of resonances in the exotic $\psi\pi$ spectrum,  which in our notation correspond to the $t$-channel.
Once we have constructed the $s$-channel amplitudes, the
$t$-channel ones can be treated similarly  (\cf Appendix~\ref{sec:our.model}).

\begin{figure}[t]
  \centering
  \includegraphics[]{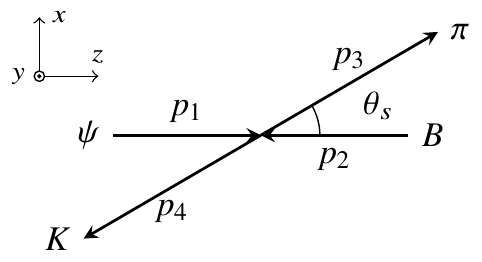}
  \caption{Scattering kinematics in the $s$-channel rest frame. In the decay kinematics, the momentum and the spin of the $\psi$ is reversed, so to keep the same helicity.}
  \label{fig:angles}
\end{figure}

In the center of mass of the $s$-channel scattering process, the $\psi$ momentum defines the $z$-axis, the momenta $p_3$ and $p_4$ lie in the $xz$-plane. We call $p$ ($q$) to the magnitude of the incoming (outgoing) three momentum.
The scattering angle $\theta_s$ is a polar angle of the pion (see Fig.~\ref{fig:angles}). The quantities depend on the Mandelstam invariants through
\begin{equation}
  \label{eq:zs}
  z_s \equiv \cos \theta_s =\frac{s(t-u)+(m_1^2-m_2^2)(m_3^2-m_4^2)}{\lambda^{1/2}_{12} \lambda_{34}^{1/2}}
  \equiv\frac{n(s,t)}{\lambda^{1/2}_{12} \lambda_{34}^{1/2}},\qquad
  p = \frac{\lambda^{1/2}_{12}}{2\sqrt{s}}, \qquad
  q = \frac{\lambda^{1/2}_{34}}{2\sqrt{s}},
\end{equation}
with $\lambda_{ik} = \left(s -  (m_i+m_k)^2\right)\left(s -  (m_i-m_k)^2\right)$.
The function $n(s,t)$ is a polynomial in $s,t$.
To incorporate resonances in the $\pi K$ system with a certain spin $j$, we expand the amplitude in partial waves,
\begin{equation}
  \label{eq:helicity-pws.s-channel}
   \mathcal{A}_\lambda(s,t,u) = \frac{1}{4\pi}\sum_{j=|\lambda|}^\infty (2j+1) A_{\lambda}^{j}(s) \,d_{\lambda0}^j(z_s),
\end{equation}
where $A^j_{\lambda}(s)$ are the helicity partial wave amplitudes in the $s$-channel.
In Eq.~\eqref{eq:helicity-pws.s-channel} the entire $t$ dependence enters though the $d$ functions. The $d$ functions have singularities in $z_s$ which lead to kinematical singularities in $t$ of the helicity amplitudes $\mathcal{A}_\lambda$.
The dynamical singularities in $t$, related to, for example, the possible resonances in the $\psi\pi$ channel,
can only be reproduced if the the sum contains the infinite number of partial waves.
In practice the $t$- or $u$-channel resonances (singularities) are accounted for explicitly through $t$- or $u$-channel partial waves,
and to avoid double counting each series is truncated at a finite number of terms.
This defines the so-called isobar model in which
 \begin{equation}
  \label{eq:isobar.model}
  \mathcal{A}_\lambda(s,t,u) = A_\lambda^{(s)}(s,t,u) + A_\lambda^{(t)}(s,t,u) + A_\lambda^{(u)}(s,t,u),
\end{equation}
with,
\begin{equation}
\label{eq:isobar}
A_\lambda^{(s)}(s,t,u) = \frac{1}{4\pi}\sum_{j=|\lambda|}^{J_\text{max}} (2j+1) A_{\lambda}^{(s) j}(s) \,d_{\lambda0}^j(z_s),
\end{equation}
where $J_\text{max}$ is finite. The expressions for the $(t)$ and $(u)$ isobars are similar to Eq.~\eqref{eq:isobar}.
Note, that due to the superscript $(s)$ the amplitudes $A_{\lambda}^{(s) j}(s)$ are not identical to the helicity partial waves, $A_{\lambda}^{j}(s)$ of Eq.~\eqref{eq:helicity-pws.s-channel}.
This is because the other two terms on the right hand side of Eq.~\eqref{eq:isobar.model} also contribute to the $s$-channel partial wave expansion.
We refer to the former as the isobar partial waves or simply, isobars. The difference between the partial waves, which are defined in a model independent way, and isobars, which appear in the specific model as in Eq.~\eqref{eq:isobar.model},
has important consequences when establishing the relation between phases of the isobar amplitudes and those of the partial waves~\cite{Khuri:1960zz,Niecknig:2012sj,Niecknig:2015ija,Danilkin:2014cra,Guo:2015zqa,Pilloni:2016obd,Albaladejo:2017hhj}.
This issue, however, is not directly related to the topic of this paper and we do not discuss it any further.

We return to the partial wave expansion, and  proceed with the analysis of kinematical singularities.
An extensive discussion and the full characterization of these singularities can be found in~\cite{Hara:1964zza,Wang:1966zza,Jackson:1968rfn,Cohen-Tannoudji:1968kvr,Martin:1970,Collins:1977jy}.
We recall that $d_{\lambda0}^j(z_s) = \hat d_{\lambda0}^j(z_s) \xi_{\lambda 0}(z_s)$, where $\xi_{\lambda 0}(z_s) = \left(\sqrt{1-z_s^2}\right)^{|\lambda|} =\sin^{|\lambda|} \theta_s$ is the so-called half angle factor that contains  all the kinematical singularities in $t$. The reduced rotational function $\hat d_{\lambda0}^j(z_s)$ is a polynomial in $s$ and $t$ of order $j - |\lambda|$ divided by the factor $\lambda_{12}^{(j-|\lambda|)/2}\lambda_{34}^{(j-|\lambda|)/2}$.
The helicity partial waves $A_\lambda^j(s)$ have singularities in $s$. These have both  dynamical and kinematical origin. The former arise, for example, from $s$-channel resonances.
The kinematical singularities, just like the $t$-dependent kinematical singularities, arise because of external particle spin.
We explicitly isolate the kinematic factors in $s$,
and denote the kinematical singularity-free helicity partial wave amplitudes by $\hat{A}_\lambda^j(s)$.\footnote{For fermion-boson scattering, the $\hat{A}_\lambda^j(s)$ can still have a branch point at $s = 0$, as discussed in~\cite{Cohen-Tannoudji:1968kvr}.}
First, the term $(p q)^{j - |\lambda|}$ is factorized out from the helicity amplitude $A_\lambda^j(s)$. This factor is there
 to cancel the threshold and pseudothreshold singularities in $s$ that appear in $\hat d^j_{\lambda0}(z_s)$.
Second, we follow \cite{Collins:1977jy} and introduce the additional   kinematic factor $K_{\lambda 0}$
(`$\pm$' is short for $\lambda=\pm 1$). These factors are required to account for a mismatch between the $j$ and $L$ dependence in the angular momentum barrier factors in presence of particles with spin.
Finally, the kinematical singularity-free helicity partial wave amplitudes $\hat{A}_\lambda^j(s)$ are defined by
\begin{subequations}
\begin{align}
A_0^j(s) &= K_{00}\, \left(p q\right)^j \,\hat{A}_0^j(s)\qquad \text{for } j\ge1,\\
A_\pm^j(s) &= K_{\pm0}\, \left(p q\right)^{j-1} \,\hat{A}_\pm^j(s)\qquad \text{for } j\ge1,\\
A_0^0(s) &= \frac{1}{K_{00}}\hat{A}_0^0(s)\qquad \text{for } j=0,
\end{align}
with $K_{00}$ and $K_{\pm 0}$ given by
\begin{align}
K_{00} &= \frac{m_1}{p\sqrt{s}} = \frac{2m_1}{\lambda_{12}^{1/2}},\\
K_{\pm 0} &= q= \frac{\lambda_{34}^{1/2}}{2\sqrt{s}}.
\end{align}
\end{subequations}

Specifically, it is expected that $A_\lambda^j(s) \sim p^{L_1} q^{L_2}$ at threshold, where $L_1$ and $L_2$ are the lowest possible orbital angular momenta in the given helicity  and parity combination.
This explains why $j = 0$ requires a special treatment~\cite{Martin:1970}, since for $j \ge 1$ we have $L_1 = j - 1$, but for $j = 0$ the lowest is $L_1 = j + 1$.
In addition, the $K$-factors have powers of $\sqrt{s}$ as required to ensure factorization of the vertices of Regge poles.
 Similarly, as explained before, $m_1$ is dynamical and thus the kinematical singularity-free amplitudes are not expected to contain singularities in $m_1^2$, and as will be seen below, the $m_1$ dependence of the $K$-factor takes care of that.
The $\hat A_\lambda^j(s)$ are left as dynamic functions, which are unknown in general and cannot be calculated from first principles.
Usually they are parameterized in terms of a sum of Breit-Wigner amplitudes with Blatt-Weisskopf barrier factors.

We now seek a representation of $\mathcal{A}_\lambda(s,t)$ in terms of the scalar functions, as discussed in Sec.~\ref{sec:intro}.
For the PV amplitude this is given by
\begin{equation}
A_\lambda(s,t) = \epsilon_\mu(\lambda,p_1) \left[
(p_3 - p_4)^\mu - \frac{m_3^2-m_4^2}{s} (p_3+p_4)^\mu
\right] C(s,t)
+ \epsilon_\mu(\lambda,p_1) (p_3 + p_4)^\mu B(s,t).\label{eq:cov}
\end{equation}
Although the second term in brackets may look like an extra $1/s$ pole, it cancels when multiplied by $(p_3+p_4)^\mu$.
This choice simplifies the final expressions, but we remark that any other choice of independent tensor structures would lead to the same results.
In the $s$-channel center of mass frame the  $\psi$ polarization vectors are given by
$\epsilon^\mu(\pm,p_1) = (0,\mp 1, -i, 0)/\sqrt{2}$ for the transverse polarizations
and
$\epsilon^\mu(0,p_1) = (|p_1|/m_1, 0, 0, E_1/m_1)$ for the longitudinal polarization.
The energies $E_i$ are calculated from the momenta and are fully determined by $s$. The functions $B(s,t)$ and $C(s,t)$ are the kinematical singularity free scalar amplitudes discussed in the Sec.~\ref{sec:intro}.

We can match Eqs.~\eqref{eq:helicity-pws.s-channel} and~\eqref{eq:cov}, and express the scalar functions as a sum over kinematical singularity free helicity partial waves.
The ratio $A_\lambda(s,t)/(K_{\lambda0}\,\xi_{\lambda 0}(z_s))$, computed using Eq.~\eqref{eq:cov}, is compared to the same ratios computed using the helicity partial waves from Eq.~\eqref{eq:helicity-pws.s-channel}. This yields
\begin{align}
  \label{eq:matching.B}
  -C(s,t) \frac{n(s,t)(s+m_1^2-m_2^2)}{4m_1^2s} + B(s,t) \frac{\lambda_{12}}{4m_1^2}
  &= \frac{A_0(s)}{K_{00}\,\xi_{0 0}(z_s)} =
  \frac{1}{4\pi} \left(
  \sum_{j>0} (2j+1) (pq)^{j}\hat{A}_{0}^{j}(s) \hat{d}_{00}^j(z_s) + \frac{\lambda_{12}}{4m_1^2}\hat{A}_{0}^{0}(s)
  \right),\\
  \label{eq:matching.C}
  \pm\sqrt{2} C(s,t) &= \frac{A_\pm(s)}{K_{\pm0}\,\xi_{1 0}(z_s)} =
  \pm\frac{1}{4\pi} \sum_{j>0} (2j+1) (pq)^{j-1}\hat{A}_{\pm}^{j}(s)\, \hat{d}_{10}^j(z_s),
\end{align}
from $\lambda = \pm$ and  $\lambda=0$, respectively, which can  be combined to
obtain
\begin{equation}
    \label{eq:B}
  4\pi B(s,t) = \hat{A}_{0}^{0}(s) + \frac{4m_1^2}{\lambda_{12}} \sum_{j>0} (2j+1) (pq)^{j} \left[\hat{A}_{0}^{j}(s) \hat{d}_{00}^j(z_s) +
  \frac{s+m_1^2-m_2^2}{\sqrt{2}m_1^2} \hat{A}_{+}^{j}(s)\, z_s\hat{d}_{10}^j(z_s)
  \right].
\end{equation}

Neither $B(s,t)$ nor $C(s,t)$ can have kinematical singularities in $s$ or $t$. In  Eqs.~\eqref{eq:matching.B}-\eqref{eq:B}, $\hat d^j_{10}(z_s)$ is regular in $t$, and the $s$ singularities at (pseudo)thresholds are canceled by the factor $(pq)^{j-1}$.
The latter factor contains a high-order pole at $s=0$. Such pole
is a feature of the dynamical model, and specifically arises because at $s=0$ the little group is not $SO(3)$ anymore. The latter motivates the partial wave expansion, thus it is not surprising that the truncation of the partial wave series results in such singularities~\cite{Toller1965,Collins:1977jy}.\footnote{For example, in Regge theory these poles are canceled by the  daughter Regge trajectories~\cite{Freedman:1967zz,Collins:1977jy}.} This construction hence does not constrain the poles at $s=0$.

For the same reason  the sum in Eq.~\eqref{eq:B} has no kinematical singularities in $s$ and $t$, however the $1/\lambda_{12}$ factor in front of the sum generates two poles at $s_{\pm} = (m_1\pm m_2)^2$, unless the expression in brackets vanishes at those points. This means that the $\hat A^j_\lambda(s)$ with different $\lambda$ cannot be  independent functions at  the (pseudo)threshold. Explicitly, in the limit
$s\to s_{\pm}$ at fixed $t$  one has  $z_s \to \infty$ and using~\cite{Jackson:1968rfn},
\begin{equation}
\hat d_{\lambda0}^j(z_s) \xrightarrow{z_s \to \infty} (-1)^{\frac{\lambda + |\lambda|}{2}}\frac{  (2J)!  \left[ J(2J - 1)  \right]^{1/2} }{2^J J \left[ (1 + \lambda)! (1 - \lambda)!  \right]^{1/2}} \frac{z_s^{J-|\lambda|}}{\langle j-1,0; 1, \lambda| j, \lambda\rangle}\qquad\text{for }|\lambda|\le 1,
\end{equation}one finds that the expression within the brackets in Eq.~\eqref{eq:B} behaves as
\begin{equation}
\hat{A}_{0}^{j}(s) \frac{(z_s)^j}{\langle j-1,0; 1, 0| j, 0\rangle} -
  \frac{s+m_1^2-m_2^2}{\sqrt{2}\: m_1^2} \hat{A}_{+}^{j}(s)\, \frac{(z_s)^j}{\sqrt{2} \: \langle j-1,0; 1, 1| j, 1\rangle}.
\end{equation}

This combination has to vanish to cancel the $1/\lambda_{12}$, thus one finds
(for $j>0$)
\begin{subequations}
\label{eq:general.form}
\begin{align}
\hat{A}_{+}^j(s) &= \langle j-1,0; 1, 1| j, 1\rangle \: g_j(s) + \lambda_{12}\: f_j(s),\\
 \hat{A}_{0}^j(s) &= \langle j-1,0; 1, 0| j, 0\rangle \frac{s+m_1^2-m_2^2}{2m_1^2} \: g'_j(s) + \lambda_{12} \: f_j'(s),
\end{align}
\end{subequations}
where $g_j(s)$, $f_j(s)$, $g'_j(s)$, and $f'_j(s)$ are regular functions at $s=s_\pm$, and $g_j(s_\pm)=g'_j(s_\pm)$.
Note that, while the functional form considered in Eq.~\eqref{eq:general.form} complies with the general requirements we are imposing, it actually implements more freedom than required by the former. For instance, one could take $f_j(s)=0$ without any loss of generality. The particular choice taken in Eq.~\eqref{eq:general.form}, however, turns out to be useful for the comparisons with other parameterizations (\ls and \cpt) which we will discuss  in Sec.~\ref{sec:comparison}.
Together with Eq.~\eqref{eq:general.form}, the expressions in Eqs.~\eqref{eq:cov}, \eqref{eq:matching.C} and \eqref{eq:B} provide the most general parameterization of the amplitude that incorporates the minimal kinematic dependence that generates the correct kinematical singularities as required by analyticity.

Upon restoration of the kinematic factors, the original helicity partial wave amplitudes read ($j>0$)
\begin{subequations}
\label{eq:helicitypw}
\begin{align}
A_{+}^j(s) &=  p^{j-1} q^j \bigg[\langle j-1,0; 1, 1| j, 1\rangle \: g_j(s) + \lambda_{12} \: f_j(s)\bigg], \\
A_{0}^j(s) &=  p^{j-1} q^j \bigg[\langle j-1,0; 1, 0| j, 0\rangle \: \frac{s+m_1^2-m_2^2}{2m_1\sqrt{s}}\: g'_j(s) + \frac{m_1}{\sqrt{s}} \lambda_{12}\: f_j'(s)\bigg],
\end{align}
\end{subequations}
and $A_{0}^0(s) = \lambda_{12}^{1/2}/(2m_1)\,\hat A_{0}^0(s)$, where $\hat A_{0}^0(s)$ is regular at (pseudo)threshold.
A particular choice of the functions $g_j(s)$, $g_j'(s)$, $f_j(s)$ and $f_j'(s)$ constitutes
a given hadronic model.
A specific example is given in Appendix~\ref{sec:our.model}.

\subsection{Implications for the \ls partial wave amplitudes}
The advantage of the \ls  basis is that the identification of the correct threshold factors is straightforward.
For a given system of two particles with spins $j_1$, $j_2$ and corresponding helicities $\lambda_1$, $\lambda_2$,
the relation between a two-particle state in the helicity and \ls basis is
\begin{equation}
\ket{j\Lambda;LS} = \sqrt{\frac{2L+1}{2j+1}} \sum_{\lambda_1\lambda_2}
\braket{L,0;S,\lambda_1-\lambda_2}{j\Lambda}\braket{j_1,\lambda_1;j_2,-\lambda_2}{S,\lambda_1-\lambda_2} \,\ket{j\Lambda;\lambda_1\lambda_2},
\end{equation}
where $\Lambda$ is the projection of the total angular momentum $j$.
For the $B \to \psi \pi K$ amplitude, it implies the following relation between the \ls amplitudes $G$ and the helicity amplitudes,
\begin{equation} \label{eq:ls.projection}
G_L^j(s) = \sqrt{\frac{2L+1}{2j+1}} \sum_{\lambda} \braket{L,0;1,\lambda}{j\lambda} A_\lambda^j(s).
\end{equation}

The amplitudes with $L=j\pm 1$ and $L=j$ differ by parity.
Equation~\eqref{eq:ls.projection} can be inverted to relate the helicity partial wave amplitudes with the \ls amplitudes $G_l^j(s)$,
\begin{equation}
\label{eq:from.spin.orbit}
A_{\lambda}^j(s) =  p^{j-1} q^j \left( \sqrt{\frac{2j-1}{2j+1}}\langle j-1,0; 1, \lambda| j, \lambda\rangle \hat{G}_{j-1}^{j}(s_{}) +
\sqrt{\frac{2j+3}{2j+1}} \langle j+1,0; 1, \lambda| j, \lambda\rangle p^2\hat{G}_{j+1}^{j}(s_{})\right).
\end{equation}

In Eq.~\eqref{eq:from.spin.orbit} we denoted the \ls partial wave amplitudes with the threshold factors explicitly factored out
by $\hat{G}_l^j(s)$, \ie $G_{j\pm1}^{j}(s) = p^{j\pm 1} \,q^j \,\hat{G}_{j \pm 1}^{j}(s)$.
We now compare the general expression for the helicity partial waves
 with the spin-orbit \ls partial waves.
We find that Eq.~\eqref{eq:from.spin.orbit} matches the general form in Eq.~\eqref{eq:general.form} when
\begin{subequations}
\label{eq:lsmatching}
\begin{align}
g_j(s) &= \sqrt{\frac{2j-1}{2j+1}}  \hat{G}^j_{j-1}(s), \\
f_j(s) &= \frac{1}{4s} \sqrt{\frac{2j+3}{2j+1}} \langle j+1,0; 1, 1| j, 1\rangle \: \hat{G}^j_{j+1}(s), \label{eq:fj_s_pole}\\
g^\prime_j(s) &= \frac{2m_1\sqrt{s}}{s+m_1^2-m_2^2}
\sqrt{\frac{2j-1}{2j+1}}\hat{G}^j_{j-1}(s),\\
f_j'(s) &= \frac{1}{4m_1\sqrt{s} } \sqrt{\frac{2j+3}{2j+1}} \langle j+1,0; 1, 0| j, 0\rangle \: \hat{G}^j_{j+1}(s).
\end{align}
\end{subequations}

The common lore is that the \ls formalism is intrinsically nonrelativistic. However, the matching in Eq.~\eqref{eq:lsmatching} proves that the formalism is fully relativistic, but care should be taken when choosing a parameterization of the \ls amplitude so that the expressions in Eqs.~\eqref{eq:lsmatching} are free from kinematical singularities. For example, if one  takes
 the functions $\hat G^j_{j-1}(s)$ and $\hat G^j_{j+1}(s)$  to be proportional to
Breit-Wigner functions with constant couplings, the amplitudes $g'_j(s)$ and $f'_j(s)$ would end up having a pole at $s = m_2^2 - m^2_1$, and/or a branch point at $s=0$ unexpected for boson-boson scattering. On the other hand, as discussed in Sec.~\ref{sec:s.channel},
the pole at $s=0$ is part of the dynamical model.
It is clear that using  Breit-Wigner parameterizations, or any other model for helicity amplitudes, \ie the left-hand sides of Eq.~\eqref{eq:lsmatching}, instead of the \ls amplitudes helps prevent unwanted singularities.

\section{Comparison with the Covariant Projection Method}
\label{sec:comparison}

We consider now the \cpt approach of~\cite{Chung:1993da,Chung:2007nn,Filippini:1995yc,Anisovich:2006bc}. As said, the method is based on the construction of explicitly covariant expressions. To describe the decay $a \to b c$, we first consider the polarization tensor of each particle with index $i$ and spin $j_i$, $\epsilon^i_{\mu_1\dots \mu_{j_i}}(p_i)$. Using the decay momentum $p_{bc}=(p_b-p_c)/2$ and the total momentum $P_{bc}=p_b+p_c$, we build a tensor $X_{\mu_1\dots \mu_{L}}(p_{bc},P_{bc})$ to represent the orbital angular momentum of the $bc$ system.
In order to find total angular momentum tensor, we first combine the polarizations of $b$ and $c$ into a ``total spin''  tensor $S_{\mu_1\dots \mu_{S}}\!\left(\epsilon_b,\epsilon_c\right)$
(orthogonal to the momentum $p_b+p_c$).
Then, we combine the tensor $S_{\mu_1\dots \mu_{S}}\!\left(\epsilon_b,\epsilon_c\right)$ with the orbital tensor $X_{\mu_1\dots \mu_{L}}$
and finally contract the result with the polarization of $a$, thus mimicking the \ls construction.
The tensors $S$ and $X$ have definite spin and parity, \ie are in an irreducible representation of the rotation group in the particle rest frame.
Thus they must be symmetric, traceless, and orthogonal to the total momentum in the particles system, $p_b+p_c$.
If one of the daughters is unstable, we can implement its decay in a similar way. The procedure is recursive, and relatively simple for low spins.
Together with the explicit covariance, it makes the formalism very attractive.

We use the \cpt to build the amplitude for  $B\to \psi \pi K$.
The construction of an amplitude for an arbitrary spin of the intermediate state
is cumbersome, and we limit ourselves to the special case of an intermediate $K^*$ with $j=1$. We start with the tensor amplitude for the scattering process $\psi B \to K^* \to \pi K$.
The orbital angular momentum of the decay $K^* \to \pi K$ in {\pw}wave is given by $X_\rho(q, P)$.
The tensor is constructed from a four-vector of the relative momentum $q^\mu = (p_3^\mu-p_4^\mu)/2$ and the total momentum of the system $P^\mu = p_3^\mu+p_4^\mu = p_1^\mu+p_2^\mu$.
For the PV amplitudes, the initial process $\psi B\to K^*$ is described by two waves. The corresponding orbital tensors are the unit rank-$0$ tensor for the {\sw}wave and rank-2 tensor $X_{\rho\mu}(p,P)$, with $p^\rho = (p_{1}^\rho - p_{2}^\rho)/2$, for the {\dw}wave. Hence
\begin{equation}
  \label{eq:cov.LS}
  A_\lambda(s,t) = \epsilon_\mu(\lambda,p_1) \left(-g^{\mu\nu} + \frac{P^\mu P^\nu}{s}\right)X_\nu(q,P) g_S(s) + \epsilon^\rho(\lambda,p_1) X_{\rho\mu}(p,P) \left(-g^{\mu\nu} + \frac{P^\mu P^\nu}{s}\right)X_\nu(q,P) g_D(s),
\end{equation}
where $P$ is the $K^*$ momentum. The final {\pw}wave orbital tensor is \mbox{$X_{\nu}(q,P) = q^\perp_\nu = q_\nu - P_\nu P\cdot q /s$}.
The {\dw}wave orbital tensor $X^{\rho\mu}(p,P)=3 p^\rho_\perp p^\mu_\perp/2-g^{\rho\mu}_\perp p_\perp^2/2$,
with $p_\perp^\mu = p^\mu - P^\mu \,P\cdot p/s$, and $g^{\rho\mu}_\perp = g^{\rho\mu} - P^\rho P^\mu/s$.
Explicitly,
\begin{equation}
  \label{eq:tensor.A.scatt}
  A_+(s,\theta_s) = -q \frac{\sin\theta_s}{\sqrt{2}} \left[g_S(s) + \frac{p^2}{2} g_D(s)\right], \quad
  A_0(s,\theta_s) = q \frac{E_1}{m_1} \cos\theta_s \left[g_S(s) - p^2 g_D(s) \right],
\end{equation}
and  matching with  Eq.~\eqref{eq:general.form} gives
\begin{subequations}
\begin{align}
  g_1(s) &= g_1'(s) = \frac{4\pi}{3} g_S(s),\\
  f_1(s) &= \frac{2\pi}{3s} g_D(s), \\
  f_1'(s) &= -\frac{4\pi}{3s}\frac{s+m_1^2-m_2^2}{m_1^2} g_D(s).
  \end{align}
\end{subequations}

The threshold conditions
$g_1(s_\pm) = g_1'(s_\pm)$ are satisfied, and
the functions
$f_1(s)$ and $f_1'(s)$ are regular at the thresholds.
Finally, we show the relation between the \cpt and the \ls amplitudes. The comparison with Eq.~\eqref{eq:from.spin.orbit} leads to
\begin{subequations}
\label{eq:tensor.to.ls}
\begin{align}
\frac{3}{4\pi}G_0^1(s) &=
g_S(s)\: q \: \sqrt{\frac{1}{3}}\left(\frac{E_1}{m_1}+2\right)
-g_D(s) \: q \: p^2 \: \sqrt{\frac{1}{3}}\left(\frac{E_1}{m_1}-1\right), \\
\frac{3}{4\pi} G_2^1(s) &=
g_D(s) \: q \: p^2 \: \sqrt{\frac{1}{6}}\left(2\frac{E_1}{m_1}+1\right)-g_S(s) \: q \: \sqrt{\frac{2}{3}}\left(\frac{E_1}{m_1}-1\right).
\end{align}
\end{subequations}

Although the $g_S(s)$ and $g_D(s)$ of the \cpt formalism, see Eq.~\eqref{eq:cov.LS}, are typically interpreted as the $S$ and
$D$  partial wave amplitudes, we see that this is the case
only at (pseudo)threshold $s=s_\pm$, where  the factor $E_1/m_1-1$ vanishes.
In Sec.~\ref{sec:different}  we discuss a specific example to show the numerical difference between the various approaches.

\subsection{Crossing symmetry; and the decay \texorpdfstring{$B\to \bar D \pi \pi$}{B -> Dbar pi pi}}\label{sec:BDpipi}
An issue with the \cpt formalism is the explicit violation of crossing symmetry.
The recursive procedure explained in~\cite{Chung:1993da,Chung:2007nn,Filippini:1995yc,Anisovich:2006bc} produces different scalar amplitudes if applied in the scattering or in the decay kinematics.
For the decay kinematics, the \cpt amplitude is constructed according to a chain $B\to \psi  K^*(\to K\pi)$. The tensor $X_\nu(q,P)$ describes  the {\pw}wave decay $K^*\to K\pi$ as before.
{\sw} and {\dw}waves are still possible for the decay $B\to \psi K^*$.
The same symbolic expression in Eq.~\eqref{eq:cov.LS} holds for the decay kinematics,
but $X_{\mu\nu}$ is now constructed from the relative momentum $\hat p = (P - \bar p_1)/2$ between $K^*$ and $\psi$ in the $B$ rest frame (we restored the $\bar{p}_1$ for the momentum of the $\psi$ in the decay kinematics), and orthogonalized with respect to the $B$ momentum $p_2$,
\begin{equation}
  \label{eq:cov.LS.decay}
  A_\lambda (s,t)= \epsilon_\mu^*(\lambda,\bar p_1) \left(-g^{\mu\nu} + \frac{P^\mu P^\nu}{s}\right)X_\nu(q,P) g_S(s) + \epsilon^{\rho*}(\lambda,\bar p_1) X_{\rho\mu}(\hat p,p_2) \left(-g^{\mu\nu} + \frac{P^\mu P^\nu}{s}\right)X_\nu(q,P) g_D(s),
\end{equation}
where
$X^{\rho\mu}$ depends on $\hat p_\perp^\mu = \hat p^\mu - p_{2}^{\mu}\, p_2\cdot \hat p/m_2^2$,
and $\hat g_{\perp}^{\rho\mu} = g^{\rho\mu}-p_2^\rho p_2^\mu /m_2^2$.
As mentioned before, crossing symmetry requires the helicity amplitude $A_\lambda(s,t)$ to be the same up to a phase for the decay and the scattering process.
The expressions for the helicity amplitudes read
\begin{subequations}
  \label{eq:tensor.A.decay}
  \begin{align}
  A_+(s,\theta_s) &= -q\frac{\sin\theta_s}{\sqrt{2}}\left( g_S(s) + p^2g_D(s)\frac{s}{2m_2^2} \right), \\
  A_0(s,\theta_s) &= q\cos\theta_s \left(\frac{E_1}{m_1}g_S(s) - \gamma(s) \: p^2 \: g_D(s)\frac{s}{m_2^2}\frac{s-m_1^2-m_2^2}{2m_1m_2} \right),
  \end{align}
\end{subequations}
where $\gamma(s)=(s-m_1^2+m_2^2)/(2 m_2 \sqrt{s})$ is the boost factor of $K^*$ in the $B$ rest frame discussed in~\cite{Filippini:1995yc, Chung:1993da}.
The matching to the general form in Eq.~\eqref{eq:general.form} is analogous. Although Eqs.~\eqref{eq:tensor.A.scatt} and~\eqref{eq:tensor.A.decay} agree at
threshold, the dynamical models differ in general, and the additional factors appearing in Eq.~\eqref{eq:tensor.A.decay} are part of the model.

The issue with the crossing symmetry is particularly interesting, and we want to illustrate it further on a simpler example. We consider the decay $B\to \bar D\pi\pi$. This reaction has been analyzed by LHCb using the \cpt formalism~\cite{Aaij:2015sqa}. Since none of the external particles have spin, the reaction is described by a single scalar amplitude.
Because of crossing symmetry, the amplitude is the same scalar function of the Mandelstam variables in  both scattering and decay kinematics.
We consider the $s$-channel scattering kinematics $B D \to \pi^+\pi^-$. We use the indices $1$, $2$, $3$ and $4$ for the $D$, $B$, $\pi^+$ and $\pi^-$ momenta. Therefore, $P = p_1 + p_2$ is the center of mass momentum, $s = P^2$ is the invariant mass of the $\pi\pi$ system, and $p$ and $q$ are the breakup momenta for the initial and final states. For simplicity, we restrict this discussion to the case of a spin-$1$ isobar, $B D \to \rho \to \pi^+\pi^-$.
The \cpt amplitude is given by
\begin{equation}
\label{eq:Dpipi.proj.tensor.ampl}
\mathcal{A}(s,t) = X_\mu(p,P) \left(-g^{\mu\nu}+\frac{P^\mu P^\nu}{s}\right) X_\nu(q,P) \: g_P(s)/
\end{equation}
In the center of mass frame of the $s$-channel, $X^\mu(p,P) = \left(0,0,0,p\right)$ and $X^\nu{(q,P)} = \left(0,q\sin \theta_s,0,q \cos\theta_s\right)$ are purely spacelike vectors proportional to the breakup momenta. Therefore,
the amplitude in Eq.~\eqref{eq:Dpipi.proj.tensor.ampl} matches the expectations.
For the decay process, the orbital tensor $X^\mu(p,P)$ is replaced by $X^\mu(\hat{p},p_2)$
to be orthogonalized to
the four-momentum $p_2$. As a result, a factor $\gamma(s) = (s-m_1^2+m_2^2)/(2m_2\sqrt{s})$ appears, and the breakup momentum from the $B \to \bar D \rho$ orbital tensor is evaluated in the rest frame of $B$.
The amplitudes for the decay process crossed to the scattering kinematics are
\begin{equation}
\label{eq:Dpipi.results}
\mathcal{A}_{[BD\to\pi \pi]} = p q \cos\theta_s\, g_P(s),\quad
\mathcal{A}_{[B\to \bar D\pi\pi]} = \gamma(s) \frac{\sqrt{s}}{m_2} p q \cos\theta_s\, g_P(s).
\end{equation}
The two amplitudes differ by a factor $\gamma(s)\sqrt{s}/m_2=(s-m_1^2+m_2^2)/(2m_2^2)$.
While this factor is analytic in $s$ and does not spoil the counting of kinematical singularities discussed in the previous section,
its appearance breaks crossing symmetry and this shows the drawback of the \cpt formalism.

The issues arise from the construction of an amplitude as subsequent  one-to-two decays.
At first sight this appears as a natural choice. However,
a well defined amplitude should have only asymptotic states on the external legs. This would exclude any decay into a resonance.
One needs to take a step back to the definition of a resonance, \ie a pole in the scattering amplitude.
Therefore, the consistent procedure would be to write the amplitude in the scattering kinematics and then
use crossing symmetry to analytically continue the amplitude into the decay region.

\subsection{\texorpdfstring{$K\pi$}{K pi}-mass distribution in different approaches}
\label{sec:different}

To explore the differences between the various approaches,
we consider the example of two intermediate vectors in the $\pi K$ channel: the $K^*(892)$, with mass and width $M_{K^*}=892$~MeV, $\Gamma_{K^*} = 50$~MeV, and the  $K^*(1410)$, with $M_{K^*}=1414$~MeV, $\Gamma_{K^*} = 232$~MeV.
The differential width is given by the  expression,
\begin{equation}
\frac{\diff \Gamma}{\diff s} = \sum_j N_j\left(\left|A_{0}^j(s)\right|^2+2\left|A_{+}^j(s)\right|^2\right)\rho(s),
\end{equation}
where $\rho(s) = \lambda_{12}^{1/2}\lambda_{34}^{1/2}/s$, and $N_j$ is a normalization constant.
\begin{figure}
\centering
\includegraphics[width=0.48\textwidth]{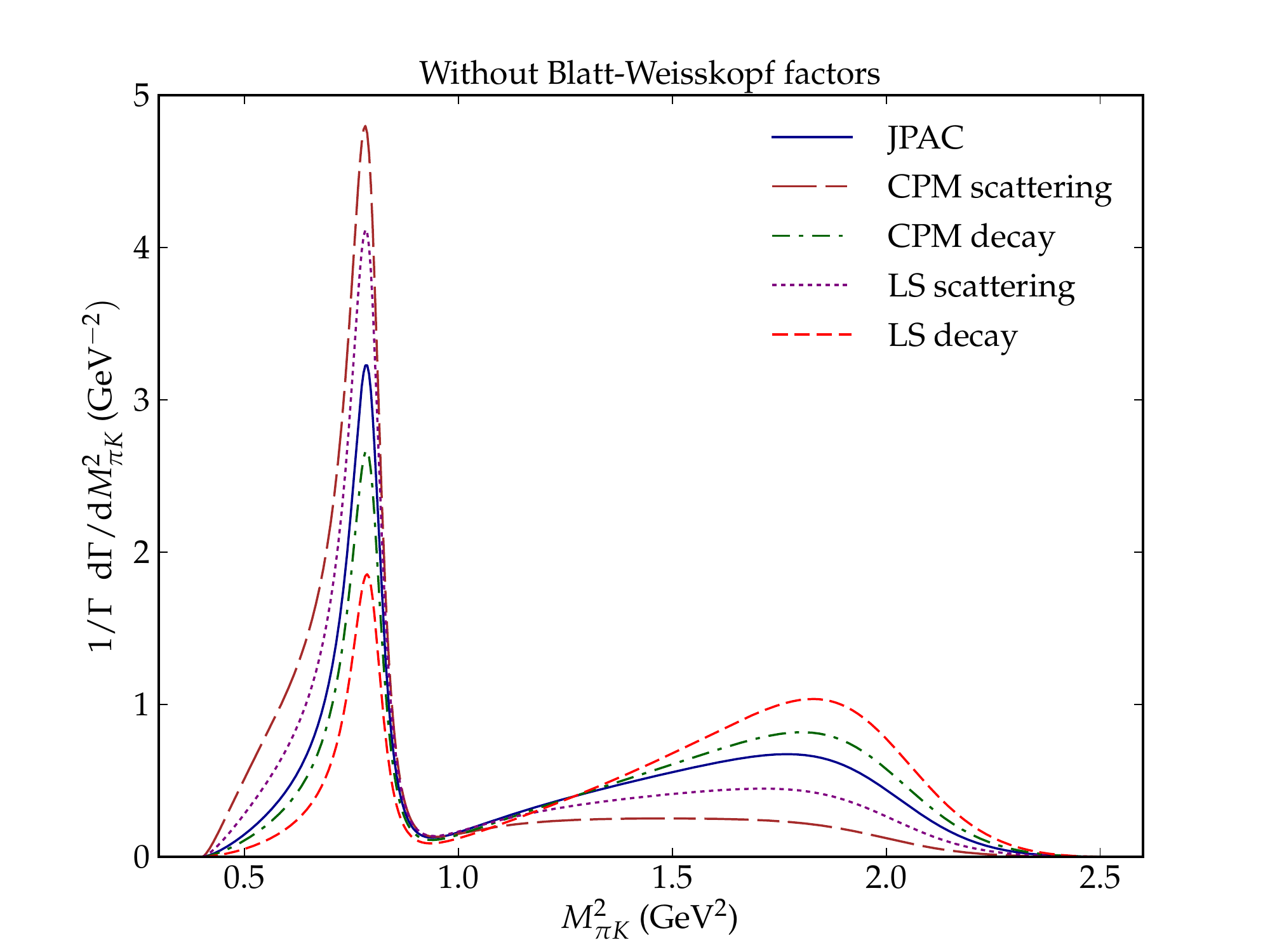}\quad
\includegraphics[width=0.48\textwidth]{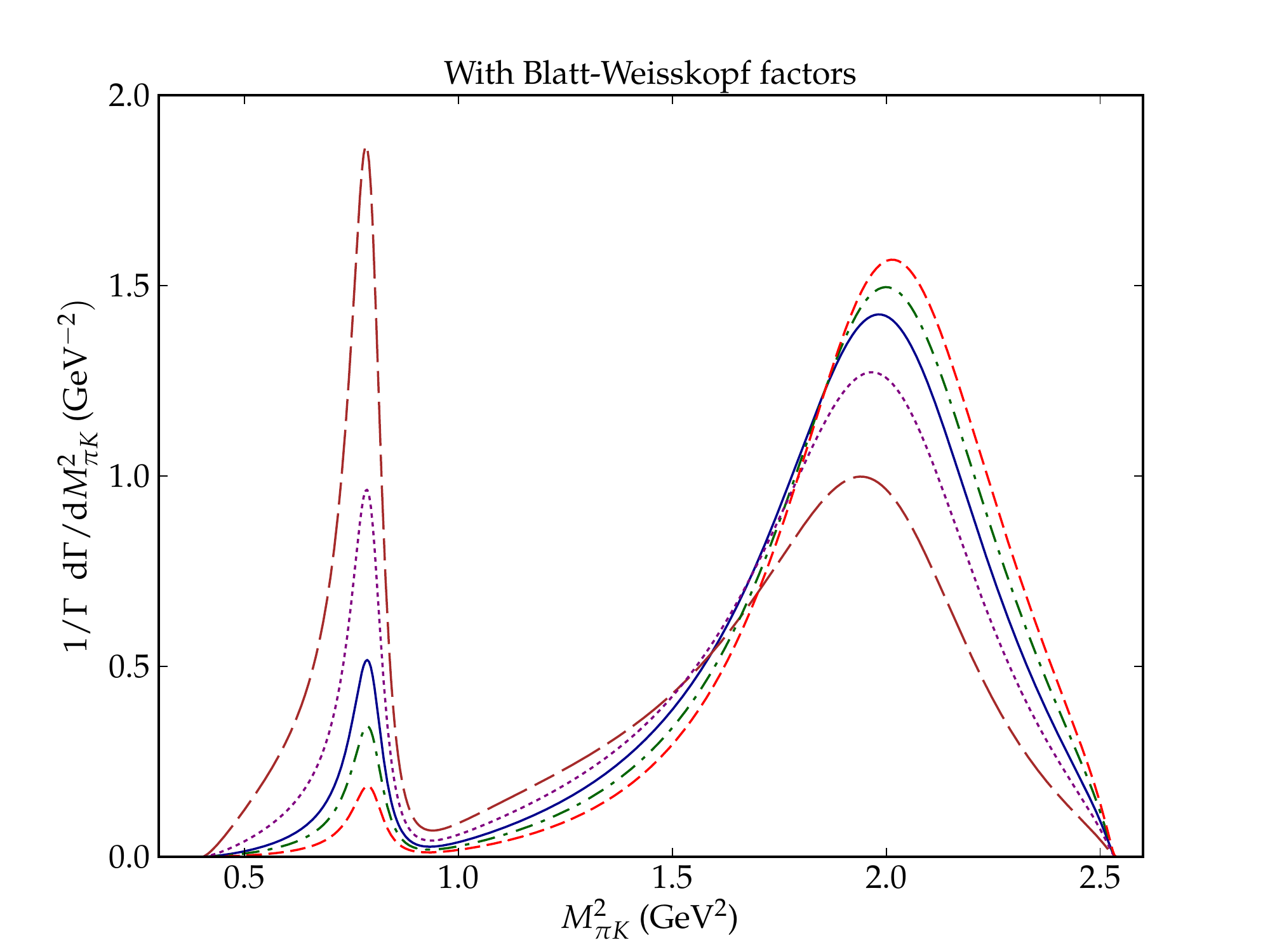}
\caption{Comparison of the lineshape of $K^*(892)$ and $K^*(1410)$ in the $\pi K$-invariant mass distribution, constructed with the different formalisms.
In the left panel we show the result with no barrier factors.
In the right panel, we include the customary Blatt-Weisskopf factors.}
\label{fig:K1410.three.methods}
\end{figure}
In Fig.~\ref{fig:K1410.three.methods} we show the results for five different scenarios.
We consider the \cpt formalisms discussed in Eq.~\eqref{eq:tensor.A.scatt} and Eq.~\eqref{eq:tensor.A.decay} (for the scattering and decay kinematics, respectively), setting $g_S(s)=0$  and $g_D(s) = T_{K^*}(s)$, with
\begin{equation}
T_{K^*}(s) \equiv \frac{0.1}{M_{K^*(892)}^2 - s - i M_{K^*(892)} \Gamma_{K^*(892)}} + \frac{1}{M_{K^*(1410)}^2 - s - i M_{K^*(1410)} \Gamma_{K^*(1410)}}.
\end{equation}
For the \ls formalism, we choose the couplings in Eq.~\eqref{eq:from.spin.orbit} to be $\hat{G}_0^1(s)=0$,  $\hat{G}_2^1(s)=T_{K^*}(s)$. The \ls amplitude in the decay kinematics differs from the one in the scattering kinematics only because of the breakup momentum of $B \to \psi K^*$, calculated in the $B$ rest frame or in the $K^*$ rest frame, respectively.
Finally, we draw a line for our proposal given by Eq.~\eqref{eq:sing.free.hatA}, the only nonzero term in the sum is $F_2^1(s) = T_{K^*}(s)$.

The partial wave amplitudes for two-to-two scattering processes are proportional to $p^{L_1}q^{L_2}$, where $p$ ($q$) are the initial (final) state break up momentum and the particles  in $L_1$($L_2$)-waves. This behavior comes from the expansion of the amplitude at threshold, and generates an unphysical growth at higher energies.
This behavior is customarily modified by
model-dependent form factors.\footnote{\ie having left-hand singularities only.}
The most popular approach is based on a nonrelativistic model introduced by Blatt and Weisskopf~\cite{Blatt:1952ije,VonHippel:1972fg}.
For the plot on the right in Fig.~\ref{fig:K1410.three.methods} we multiply the amplitudes by the Blatt-Weisskopf barrier factors
\begin{subequations}
\begin{align}
  B_1(q) &= \sqrt{\frac{1}{1+q^2R^2}}, \\
  B_2(p) &= \sqrt{\frac{1}{9+3p^2R^2+p^4R^4}},
\end{align}
\end{subequations}
for the initial {\pw} and final {\dw}waves, respectively.
The couplings are set as $g_S(s) = \hat{G}_0^1(s) = F_0^1(s) = 0$ and $g_D(s) = \hat{G}_2^1(s) = F_2^1(s) = T_{K^*}(s)B_1(q)B_2(p)$ for the corresponding formalisms in Eqs.~\eqref{eq:cov.LS}, \eqref{eq:tensor.A.scatt} and \eqref{eq:sing.free.hatA}.
The constant $R$ is chosen to be $5\,$GeV$^{-1}$, which corresponds $1\,$fm, \ie the scale of the strong interaction.

We see in Fig.~\ref{fig:K1410.three.methods} that the $K\pi$ invariant mass squared distribution is distorted differently in all models.
It is straightforward to track down where the differences come from. In the JPAC amplitude
 of Eqs.~\eqref{eq:sing.free.hatA} and~\eqref{eq:sing.free.hatA2}, the threshold factor in the $F_2^1(s)$ function in Eq.~\eqref{eq:sing.free.hatA} is set to $\lambda_{12}$,
in contrast to the \cpt and \ls formalisms where the factor $p^2 = \lambda_{12}/(4s)$ is used.  This makes the differential width distribution different by the factor $1/s^2$. Another difference originates from the  factor $E_1/m_1$ for the $A_0(s)$ amplitude which was required by analyticity. We showed that it is not present in the \ls approach, as one can see in Eq.~\eqref{eq:from.spin.orbit}.
In the physical domain this factor behaves as $1/\sqrt{s}$ at the amplitude level, resulting in $1/s$ difference in the  the differential width.
\section{Summary and Conclusions}
\label{sec:sc}
We considered different approaches for constructing amplitudes for scattering and decay processes.
Although the problem might be viewed as standard exercise, there seems to be confusion  among amplitude analysis practitioners, as to which formalism best 
represents $S$-matrix constraints~\cite{Chung:2007nn,Filippini:1995yc}. Specifically, we have compared the
canonical helicity formalism~\cite{Jacob:1959at,Chung:1971ri,Collins:1977jy,cookbook} and
the covariant projection method~\cite{Chung:1993da,Chung:2007nn,Filippini:1995yc,Anisovich:2006bc}.
We used analyticity as a guiding principle to examine these approaches.
Using as example the decay $B\to \psi \pi K$, and the helicity formalism, we separated the kinematical factors from the dynamical functions. We then  matched the helicity amplitudes with  the most general covariant expression.
In this process  we identified  kinematical constraints on the helicity amplitudes.
We have shown that the  na\"ive parameterization of the \ls couplings fails to satisfy all the constraints required by analyticity.
We found that, in contrast to \ls parameterization, an extra factor \mbox{$(s+m_1^2-m_2^2)/(2m_1\sqrt{s})$} naturally appears in the tensor formalism when written for the scattering kinematics.
More interestingly, the customary recipes in the \cpt approach  explicitly violate crossing symmetry.
In particular, we showed that the tensor approach discussed in~\cite{Chung:1993da,Chung:2007nn,Filippini:1995yc,Anisovich:2006bc},
when applied to the decay kinematics directly, introduces a peculiar energy dependence which has no clear physical motivation.

To address the issue of the relativistic corrections, we recall the relation between the helicity and the \ls amplitudes.
This relation is valid for any energy.
The concept of the spin-orbit decomposition is fully relativistic.
However, analyticity prevents the \ls couplings to be parameterized as simple constants.
We remark that our observations and conclusions are strictly valid only when asymptotic states are considered.
We performed extensive studies for the four-legs process which describes two-to-two scattering, or
the one-to-three decay, when the mother particle has an (infinitely) narrow width.
The extension to other reactions requires dedicated studies.

\begin{acknowledgments}
We are grateful for the inspiring atmosphere at the
PWA9/ATHOS4 workshop where the idea for this work was born.
AP thanks Greig Cowan and Jonas Rademacker for useful discussions about the application of the different formalisms in LHCb.
MM and AP would like to thank Suh-Urk Chung and Andrey Sarantsev for several useful discussion about the \cpt approach,
Dmitry Ryabchikov for sharing his experience with partial wave analysis
using the Zemach and \ls formalisms, and Anton Ivashin for several useful discussions.
This work was supported by BMBF,
the U.S.~Department of Energy under grants No.~DE-AC05-06OR23177 and No.~DE-FG02-87ER40365,
PAPIIT-DGAPA (UNAM, Mexico) grant No.~IA101717,
CONACYT (Mexico) grant No.~251817,
Research Foundation -- Flanders (FWO),
U.S.~National Science Foundation under award numbers PHY-1507572, PHY-1415459 and PHY-1205019,
and Ministerio de Econom\'ia y Competitividad (Spain) through grant
No.~FPA2016-77313-P.
\end{acknowledgments}

\appendix

\section{Parity-conserving amplitudes in the $s$ channel}
\label{sec:parity.conserving}
We consider the most general scattering amplitude for $\psi B\to \pi K$ with parity conservation enforced.
The only tensor structure allowed is
\begin{equation} \label{eq:cov.pc.s}
A_\lambda(s,t,u) = -i\epsilon_{\mu\nu\rho\sigma} p_1^\mu p_2^\nu p_3^\rho \epsilon(\lambda,p_1)^\sigma D(s,t),
\end{equation}
where the Levi-Civita tensor is defined by $\epsilon_{0123} = 1$ and
$D(s,t)$ is the singularity free scalar amplitude.

As follows from the derivation in~\cite{Collins:1977jy},
the kinematical factor is $K_{\pm0} = \sqrt{s}\: p q=\lambda_{12}^{1/2}\lambda_{34}^{1/2}/(4\sqrt{s})$, \ie
the minimal $L$ in the final and initial states matches $j$.
Removing also the half-angle factor $\xi_{10}(z_s) =\sin\theta_s$, we get
\begin{equation}
  \label{eq:A0hat}
  \frac{A_{\pm}(s,\theta_s)}{K_{\pm 0}\: \xi_{\pm 0}(z_s)}
  =  \frac{1}{\sqrt{2}} D(s,t),
\end{equation}
where the partial wave helicity amplitudes can be written as
\begin{align}
  A_{\pm}(s,\theta_s) &= \frac{1}{4\pi} \sum_{j>0} (2j+1)\, K_{\pm0}  \: (pq)^{j-1} \hat{A}^j_{\pm}(s)\, \hat{d}_{\pm0}^j(z_s) \sin\theta_s.
\end{align}

Finally, we can match the scalar amplitude $D(s,t)$ with the  partial waves helicity amplitudes,
\begin{equation}
  \label{eq:D}
  D(s,t) = \frac{\sqrt{2}}{4\pi} \sum_{j>0} (2j+1) (pq)^{j-1}\hat{A}_{+}^{j}(s)
  \hat{d}_{+0}^j(z_s).
\end{equation}

\section{The minimal singularity-free parameterization of the helicity amplitudes}
\label{sec:our.model}
We consider a model for the reaction $B\to \psi \pi K$, using the general parameterization of the helicity amplitudes
obtained in Sec.~\ref{sec:s.channel} and Appendix~\ref{sec:parity.conserving}.
We use the isobar model of Eq.~\eqref{eq:isobar.model}, and neglect any $u$-channel ($\psi K$) resonant contribution.
We already provided the covariant form of the $s$-channel amplitude $A_\lambda^{(s)}(s,t)$ in Eq.~\eqref{eq:cov}, where the scalar functions
$B$ and $C$ are related to the kinematical singularity free helicity partial waves $\hat{A}_\lambda^{(s),j}$.
The general form is in Eq.~\eqref{eq:general.form}.
A particular choice of the functions
$g_j^{(s)}$, $g_j^{(s)\prime}$, $f_j^{(s)}$ and $f_j^{(s)\prime}$ determines our model.
To match the \ls parameterization at threshold, we set
\begin{align}
g_j^{(s)} &= \sqrt{\frac{2j-1}{2j+1}} F_{j-1}^{(s),j},&
f_j^{(s)} &= \sqrt{\frac{2j+3}{2j+1}} \langle j+1,0;1,0|j,0 \rangle F_{j+1}^{(s),j},\\
g^{(s)\prime} &= \sqrt{\frac{2j-1}{2j+1}} F_{j-1}^{(s),j}, &
f_j^{(s)\prime} &= \sqrt{\frac{2j+3}{2j+1}} \langle j+1,0;1,1|j,1 \rangle F_{j-1}^{(s),j},
\end{align}
where $F_{l}^{(s),j}$ are independent dynamical functions,
\begin{widetext}
\begin{align} \label{eq:sing.free.hatA}
\hat{A}_{0}^{(s),j} &= \frac{s+m_1^2-m_2^2}{2m_1^2}\left(
\sqrt{\frac{2j-1}{2j+1}}\langle j-1,0;1,0|j,0 \rangle\coupl{j-1}{s}{j} +\lambda_{12}\:
\sqrt{\frac{2j+3}{2j+1}}\langle j+1,0;1,0|j,0 \rangle \coupl{j+1}{s}{j}
\right),\\
\hat{A}_{+}^{(s),j} &= \sqrt{\frac{2j-1}{2j+1}}
\langle j-1,0;1,1|j,1\rangle \coupl{j-1}{s}{j} +
\lambda_{12}\: \sqrt{\frac{2j+3}{2j+1}} \langle j+1,0;1,1|j,1 \rangle \coupl{j+1}{s}{j}.\label{eq:sing.free.hatA2}
\end{align}
\end{widetext}

The amplitudes in the $t$-channel are analogous to the $s$-channel ones upon replacement of momenta and masses,
$2\leftrightarrow 3$. The corresponding dynamical functions are $F_{l}^{(t),j}$.

Combining the $s$-channel PV amplitude in Eq.~\eqref{eq:cov}, the $s$-channel PC amplitude in Eq.~\eqref{eq:cov.pc.s},
and their $t$-channel counterparts, the full amplitude reads
\begin{align}\label{eq:total.A}
  A_\lambda(s,t,u) &= \epsilon^{\mu*}(\lambda,\bar p_1) A(s,t,u)_\mu = \nonumber \\
 &\quad\,\epsilon^{\mu*}(\lambda,\bar p_1)\Bigg[ \left((p_3 - p_4)_\mu - \frac{m_3^2-m_4^2}{s} (p_3+p_4)_\mu
                            \right) C^{(s)}(s,t)  \nonumber + (p_3 + p_4)_\mu B^{(s)}(s,t) \nonumber\\
  &\qquad\qquad+ \left((p_2 - p_4)_\mu - \frac{m_2^2-m_4^2}{t} (p_2+p_4)_\mu \right)
     C^{(t)}(s,t)  +(p_2 + p_4)_\mu B^{(t)}(s,t)\nonumber \\
    &\qquad\qquad -i\epsilon_{\sigma\nu\rho\mu} \bar p_1^\sigma p_2^\nu p_3^\rho  D^{(s)}(s,t)
     -i\epsilon_{\sigma\nu\rho\mu} \bar p_1^\sigma p_3^\nu p_2^\rho  D^{(t)}(s,t)\: \Bigg],
\end{align}
with $x=s,t$, and
\begin{equation}
m_{x} = \left\{\begin{matrix} m_2 \text{ for }x=s, \\  m_3 \text{ for }x=t,\end{matrix}\right.
\qquad
\lambda_{1x} = \left\{
\begin{matrix} \lambda_{12} \text{ for }x=s,\\ \lambda_{13} \text{ for }x=t,\end{matrix}\right.
\qquad
q_{x} = \left\{\begin{matrix} \lambda_{34}^{1/2}/(2\sqrt{s}) \text{ for }x=s, \\  \lambda_{24}^{1/2}/(2\sqrt{t}) \text{ for }x=t,\end{matrix}\right.
\quad p_x = \frac{\lambda_{1x}^{1/2}}{2\sqrt{x}},
\end{equation}

\begin{align}\nonumber
  C^{(x)}(s,t) &= \frac{1}{4\pi\sqrt{2}} \sum_{j>0} \sqrt{2j+1} (p_x q_x)^{j-1}\\\label{eq:final.C}
  & \qquad \times \Bigg[\coupl{j-1}{x}{j}\sqrt{2j-1}\left\langle j-1, 0; 1,1|j,1 \right\rangle + \coupl{j+1}{x}{j}\,\lambda_{1x} \sqrt{2j+3} \left\langle j+1, 0; 1,1|j,1 \right\rangle\Bigg]\, \hat{d}_{10}^j(z_x),\\\nonumber
   B^{(x)}(s,t) &= \frac{\sqrt{3}}{4\pi} \coupl{1}{x}{0} +  \frac{1}{4\pi}\frac{4m_1^2}{\lambda_{1x}} \sum_{j>0} \sqrt{2j+1} (p_x q_x)^{j}  \frac{x+m_1^2-m_x^2}{\sqrt{2}m_1^2}\nonumber\\
   &\qquad \times\Bigg[\coupl{j-1}{x}{j}\sqrt{2j-1} \left(\frac{1}{\sqrt{2}}\left\langle j-1, 0; 1,0|j,0\right\rangle \hat d^j_{00}(z_x) + \left\langle j-1, 0; 1,1|j,1\right\rangle \hat d^j_{10}(z_x) z_x \right)\nonumber\\\label{eq:final.B}
    &\qquad\qquad+\coupl{j+1}{x}{j}\,\lambda_{1x} \sqrt{2j+3} \left(\frac{1}{\sqrt{2}}\left\langle j+1, 0; 1,0|j,0 \right\rangle \hat d^j_{00}(z_x) + \left\langle j+1, 0; 1,1|j,1 \right\rangle \hat d_{10}(z_x)z_x\right) \Bigg],\\
      D^{(x)}(s,t) &= \frac{\sqrt{2}}{4\pi} \sum_{j>0} (2j+1) (p_xq_x)^{j-1} \,\coupl{j}{x}{j}\, \hat{d}_{10}^j(z_x).
\end{align}
The decay $\psi \to \mu^+ (q_{2}) \mu^-(q_{1})$ can be attached by contracting the tensor amplitude $A_\mu$
from Eq.~\eqref{eq:total.A} with the tensor given by the fermion vector current.
\begin{equation}
\sum_{\text{spins}}\mathcal{M} = A(s,t,u)_\mu \bar u(q_1) \gamma^\mu v(q_2).
\end{equation}
The square of the leptonic tensor summed over the unobserved polarizations of the leptons yields
\begin{equation}
\label{eq:mumu.decay}
\left|\mathcal{M}\right|^2
= 4 \: A(s,t,u)_\mu \:  A(s,t,u)^*_\nu \: \left(q_1^\mu q_2^\nu + q_1^\nu q_2^\mu - \frac{m_1^2}{2}g^{\mu\nu} \right).
\end{equation}
The amplitude is linear in the dynamical functions $F_l^{(x),j}$, \ie
\begin{equation}
A_\mu = \sum_{jlx} \coupl{l}{x}{j} (Z_{l}^{xj})_\mu,
\end{equation}
where the index $j$ is the spin in the $x$-channel. The sum over $j$ goes from $0$ to $\infty$
in general, while in the considered isobar model one includes only the terms with relevant resonances. The orbital angular momentum $l$ runs over $j-1,j,j+1$.
In the special case $j = 0$, the possible values of $l$ are $0$ and $1$ only.
The functions $\coupl{l}{x}{j}$ contain the dynamical input. They include the resonance amplitude (\eg parameterized by the Breit-Wigner formula) and left hand singularities (\eg Blatt-Weisskopf barrier factors).
The $(Z_{l}^{xj})_\mu$ are kinematic functions responsible for the right angular
dependence. The square of the matrix element is a bilinear form in the dynamical functions
\begin{equation}
\label{eq:bilinear}
\left|\mathcal{M}\right|^2
= \sum_{jj'll'xx'} \coupl{l}{x}{j} \left(\coupl{l'}{x'}{j'}\right)^* V^{xjx'j'}_{ll'},
\quad V^{xjx'j'}_{ll'} = 4\left[\left(Z_{l}^{xj}\right)_\mu \left(q_1^\mu q_2^\nu + q_1^\nu q_2^\mu - \frac{g^{\mu\nu}}{2} m_1^2\right) \left(Z_{l'}^{x'j'}\right)_\nu^*\right].
\end{equation}
The expression for the kinematic function $(Z_{l}^{xj})_\mu$ is given in Eq.~\eqref{eq:kin.func.Z}.
The first two terms in Eq.~\eqref{eq:kin.func.Z} are used for
the PV decay, the last term for the PC one
\begin{equation}
\label{eq:kin.func.Z}
(Z_{l}^{xj})_\mu = C_\mu^{x} \zeta_{l}^{xj} + B_\mu^x \beta_{l}^{xj} +  D_\mu^x \delta_{l}^{xj}.
\end{equation}
The functions $C_\mu^x$, $B_\mu^x$, and $D_\mu^x$ describe the tensor structures in front of $C^{(x)}$, $B^{(x)}$, and $D^{(x)}$
in Eq.~\eqref{eq:total.A}. The functions $\zeta_{l}^{xj}$, $\beta_{l}^{xj}$ and $\delta_{l}^{xj}$ take over
the kinematic dependence of the functions $C^{(x)}$, $B^{(x)}$ and $D^{(x)}$, \ie they are factors in front of $\coupl{l}{x}{j}$

\begin{align}
C_\mu^{s} = (p_3 - p_4)_\mu - \frac{m_3^2-m_4^2}{s} (p_3+p_4)_\mu,\quad B_\mu^{s} = (p_3+p_4)_\mu,\quad
D_\mu^{s} = -i\epsilon_{\sigma\nu\rho\mu} \bar p_1^\sigma p_2^\nu p_3^\rho,\\
C_\mu^{t} = (p_2 - p_4)_\mu - \frac{m_2^2-m_4^2}{t} (p_2+p_4)_\mu,\quad B_\mu^{t} = (p_2+p_4)_\mu,\quad
D_\mu^{t} = -i\epsilon_{\sigma\nu\rho\mu} \bar p_1^\sigma p_3^\nu p_2^\rho,
\end{align}
with
\begin{align}
\zeta_{j-1}^{xj} &= \frac{(p_x q_x)^{j-1}}{4\pi\sqrt{2}} \sqrt{(2j+1)(2j-1)} \left\langle j-1, 0; 1,1|j,1 \right\rangle \hat{d}_{10}^j(z_x),\\
\zeta_{j}^{xj} &= 0,\\
\zeta_{j+1}^{xj} &= \frac{\lambda_{1x}(p_x q_x)^{j-1}}{4\pi\sqrt{2}} \sqrt{(2j+1)(2j+3)} \left\langle j+1, 0; 1,1|j,1 \right\rangle \hat{d}_{10}^j(z_x),\\[5mm]
\beta_{j-1}^{xj} &= \frac{4m_1^2(p_x q_x)^{j}}{4\pi\lambda_{1x}} \frac{x+m_1^2-m_x^2}{\sqrt{2}m_1^2}\sqrt{(2j+1)(2j-1)} \left(\frac{1}{\sqrt{2}}\left\langle j-1, 0; 1,0|j,0\right\rangle \hat d^j_{00}(z_x) + \left\langle j-1, 0; 1,1|j,1\right\rangle \hat d^j_{10}(z_x) z_x \right),\\
\beta_{j}^{xj} &= 0,\\
\beta_{j+1}^{xj} &= \frac{4m_1^2(p_x q_x)^{j}}{4\pi} \frac{x+m_1^2-m_x^2}{\sqrt{2}m_1^2}\sqrt{(2j+1)(2j+3)} \left(\frac{1}{\sqrt{2}}\left\langle j+1, 0; 1,0|j,0 \right\rangle \hat d^j_{00}(z_x) + \left\langle j+1, 0; 1,1|j,1 \right\rangle \hat d_{10}(z_x)z_x\right),\\[2mm]
\delta_{j-1}^{xj} &= 0,\\
\delta_{j}^{xj}\: &= \frac{\sqrt{2}}{4\pi} (2j+1) (p_xq_x)^{j-1} \hat{d}_{10}^j(z_x),\\
\delta_{j+1}^{xj} &= 0.
\end{align}

The special case $j = 0$ has $\zeta_{l}^{x0} = \beta_{0}^{x0} = \beta_{-1}^{x0} = \delta_{l}^{xj} = 0$ and $\beta_{1}^{x0} = \sqrt{3}/(4\pi)$.

\bibliographystyle{apsrev4-1}
\bibliography{quattro}
\end{document}